\documentclass[letterpaper]{article} 
\usepackage{aaai25}  
\usepackage{times}  
\usepackage{helvet}  
\usepackage{courier}  
\usepackage[hyphens]{url}  
\usepackage{graphicx} 
\usepackage{natbib}  
\usepackage{caption} 
\usepackage{multirow}
\usepackage{subcaption}
\usepackage{amsmath}
\usepackage{booktabs}

\title{Is Your Autonomous Vehicle Safe? Understanding the Threat of Electromagnetic Signal Injection Attacks on Traffic Scene Perception}

\author {
    Wenhao Liao\textsuperscript{\rm 1}\equalcontrib,
    Sineng Yan\textsuperscript{\rm 1}\equalcontrib,
    Youqian Zhang\textsuperscript{\rm 2}\footnote{Corresponding author.},
    Xinwei Zhai\textsuperscript{\rm 1},
    Yuanyuan Wang\textsuperscript{\rm 1},\\
    Eugene Yujun Fu\textsuperscript{\rm 3}
}
\affiliations {
    \textsuperscript{\rm 1}Shenzhen University, Shenzhen, China\\
    \textsuperscript{\rm 2}The Hong Kong Polytechnic University, Hong Kong S.A.R., China\\
    \textsuperscript{\rm 3}The Education University of Hong Kong, Hong Kong S.A.R., China\\
    liaowenhao23@email.szu.edu.cn, 
    yansineng2023@email.szu.edu.cn,
    you-qian.zhang@polyu.edu.hk,
    zhaixinwei2022@email.szu.edu.cn, 
    yywang@szu.edu.cn,
    eugenefu@eduhk.hk
    
}

\usepackage{bibentry}

\begin{document}

\maketitle

\begin{abstract}
Autonomous vehicles rely on camera-based perception systems to comprehend their driving environment and make crucial decisions, thereby ensuring vehicles to steer safely. However, a significant threat known as Electromagnetic Signal Injection Attacks (ESIA) can distort the images captured by these cameras, leading to incorrect AI decisions and potentially compromising the safety of autonomous vehicles. Despite the serious implications of ESIA, there is limited understanding of its impacts on the robustness of AI models across various and complex driving scenarios. To address this gap, our research analyzes the performance of different models under ESIA, revealing their vulnerabilities to the attacks. Moreover, due to the challenges in obtaining real-world attack data, we develop a novel ESIA simulation method and generate a simulated attack dataset for different driving scenarios. Our research provides a comprehensive simulation and evaluation framework, aiming to enhance the development of more robust AI models and secure intelligent systems, ultimately contributing to the advancement of safer and more reliable technology across various fields.
\end{abstract}

\section{Introduction}

Autonomous driving is making rapid progress, and becoming a reality with real products now being used in everyday life.
Notable examples, including Tesla, with its Autopilot, Google's Waymo, Baidu's Apollo Go, etc., exemplify the cutting-edge autonomous driving capabilities in practice~\citep{badue2021self}. 
However, safety concerns still hinder the public trust and constrain the wide adoption of autonomous driving, particularly with the potential threat of adversarial attacks~\citep{liu2020exploring, ljubi2023role, nastjuk2020drives}. Unlike human drivers, autonomous vehicles can be potentially affected by adversarial attacks, leading to inaccurate perception of driving environment, and eventually, unsafe operation~\citep{zhang2021evaluating}.

\begin{figure}[t]
\centering
\includegraphics[width=0.9\columnwidth]{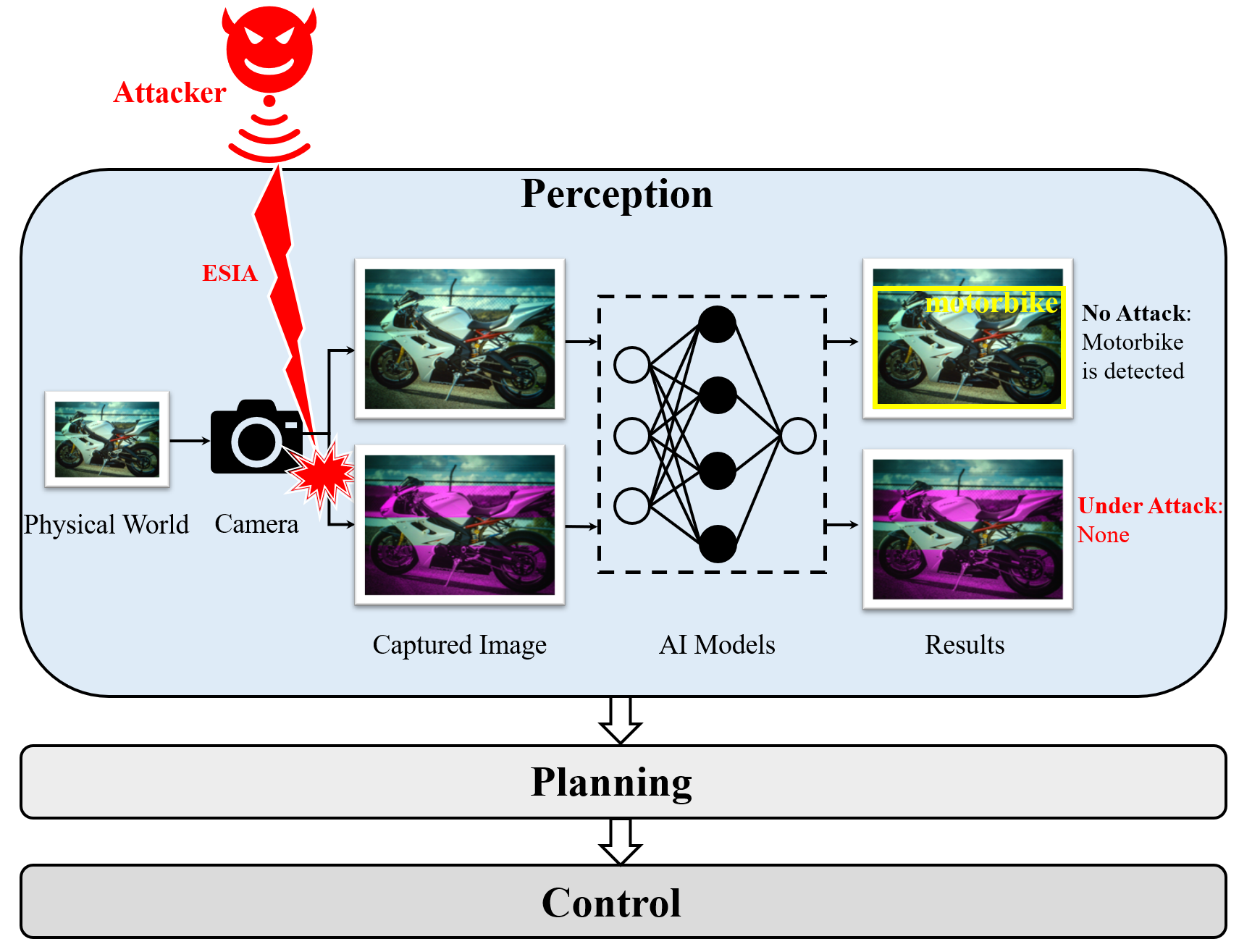} 
\caption{A general process of autonomous driving systems consists of three modules: Perception, Planning, and Control. Electromagnetic signal injection attacks (ESIA) can manipulate the output image of the autonomous vehicle's cameras, hiding the motorbike from being detected, potentially leading to an accident.}
\label{fig:high_level_structure}
\end{figure}

Figure~\ref{fig:high_level_structure} illustrates a general process of autonomous driving systems, composed of three modules: ``Perception'', ``Planning'', and ``Control''. 
The perception module gathers input data through sensors such as cameras, LiDARs, and radars, then employs artificial intelligence (AI) to accurately interpret road markings, vehicles, pedestrians, and other environmental elements for further processing. 
The planning module uses this environmental understanding to generate driving strategies, such as motion trajectories, while the control module converts these strategies into control commands to steer wheels so as to achieve autonomous driving.

The perception module is arguably the most critical, which serves as the fundamental component and directly influences the decisions made by the other modules. It is thus the primary target of adversarial attacks~\citep{guesmi2023physical}.
Of specific interest is the camera and AI models within the perception module due to their versatility and critical role, including functions such as obstacle detection, lane recognition, and traffic sign identification. If such a perception module is attacked, it may result in traffic accidents involving the autonomous car. For example, failing to detect obstacles ahead, the car may not stop, leading to a catastrophic car crash.
It is essential to point out that such attacks are not imagined, and many studies have been demonstrated to manipulate images captured by cameras through different attack vectors, including adversarial patches/stickers, lasers, ultrasound, and electromagnetic interference (detailed discussion is presented in Related Work). 

Among these attacks, one emerging and concerning threat is Electromagnetic Signal Injection Attack (ESIA)~\citep{kohler2022signal, jiang2023glitchhiker, zhang2024modeling, zhang2024understanding, kang2024anti}, a real-world attack that injects adversarial electromagnetic signals into the camera circuits, causing malicious image distortion, e.g., color strips as shown in Figure~\ref{fig:high_level_structure}.
~\citet{jiang2023glitchhiker}, and ~\citet{zhang2024understanding}~provided a detailed workflow and error modeling of ESIA, demonstrating and confirming that ESIA can cause color strips. 
~\citet{jiang2023glitchhiker} also showed that it is feasible for attackers to deploy compact attack devices, as small as a credit card, close to car cameras and trigger the attack while the car is in operation.
ESIA could mislead AI models within the autonomous driving system, leading to erroneous decisions with potentially catastrophic consequences for road safety. 
Understanding the robustness and reliability of the existing AI models under such attacks is crucial for ensuring safety, building public trust, and enhancing future models to better handle under-attack scenarios.
Although related work has explored various attacks on cameras, the focus of this paper lies on the severe implications of ESIA on the perception system and the inherent risks it poses to autonomous driving operations, which have not been investigated yet.

A critical research gap exists in understanding the varying severity of ESIA attacks across different driving scenarios. 
Uncertainties persist regarding the specific risks posed by such attacks in diverse environments, necessitating a comprehensive investigation to gauge their potential impact accurately. 
However, conducting practical experiments to gather extensive under-attack data is hindered by the complexity and cost associated with setting up professional-grade attack systems and camera systems that save images in raw format~\citep{jiang2023glitchhiker, zhang2024modeling}. This limits the research on understanding the threat posed by ESIA on real traffic images. Additionally, comprehending why AI models falter under attack conditions presents a significant challenge, complicating efforts to bolster system defenses against adversarial threats.

To bridge these research gaps, we made the following contributions:

\begin{itemize}
    \item We develop a novel simulation method that mimics ESIA's adversarial patterns, making the attack scalable and facilitating extensive research without the need for costly practical setups. 
    \item We introduce an ESIA simulation dataset focused on traffic scenarios, categorized by the attack severity, which will aid in the development of robust and reliable autonomous driving systems in the future.
    \item We gain a deep understanding of how different models perform under ESIA attacks through systematic experiments in diverse traffic scenarios, highlighting the threat of ESIA and potential safety risks. 
    
\end{itemize}

\section{Related Work}
\label{sec:related_work}

Adversarial attacks on cameras manifest in two primary forms: digital attacks, and physical attacks.

\subsection{Digital Attacks} 

In digital attacks, an attacker can arbitrarily alter the image at a granular pixel level. 
In an early work,~\citet{szegedy2014intriguing} demonstrated how even minor perturbations, imperceptible to the human eye, could significantly disrupt the performance of AI models. 
Following this, numerous algorithms, for example,~\citet{goodfellow2015explaining, carlini2017towards, chen2020hopskipjumpattack}, have been developed to craft adversarial perturbations that exploit the vulnerabilities in the model's learning process, causing it to make incorrect predictions or classifications.

\subsection{Physical Attacks} 

Physical attacks involve manipulating inputs from the physical world to deceive the camera system. Several techniques have been explored in this domain.

Adversarial patches or stickers, strategically positioned on physical surfaces of objects, aim to deceive recognition systems, leading to misclassification or false interpretations. 
For instance, a carefully designed sticker on a stop sign could cause an autonomous vehicle's camera to misinterpret it~\citep{eykholt2018robust, song2018physical, duan2020adversarial}, or an adversarial marking on the ground can cause an autonomous vehicle to drive in the wrong direction~\citep{jing2021too}.

Beyond object manipulation, other attack modalities aim to interfere with the cameras directly.
Light-based attacks aim to alter the lighting conditions around the camera, potentially causing crucial visual information to be overlooked or misinterpreted.
For example, directing a strong light source at the camera can result in temporary blindness~\citep{petit2015remote, yan2016can, fu2021remote}, or fine-grained laser pointed at the camera can achieve a malicious control of traffic light color~\citep{yan2022rolling}.
Acoustic-based manipulation involves broadcasting ultrasound to cause perturbations in the car camera's stabilizer, leading to blurred images, so that the detection is severely degraded~\citep{ji2021poltergeist, zhu2023tpatch}.
Furthermore, electromagnetic manipulation, although a significant concern, has been previously discussed and is not repeated here.

\section{Simulation of ESIA}

This section introduces a novel simulation method for generating adversarial patterns. The similarity between these simulated patterns and those produced by real-world attacks is then verified. 
Finally, the method is employed to create a dataset of simulated adversarial patterns.

\subsection{Simulation Method}

Normally, a camera transmits its captured images to subsequent processing stages as illustrated in Figure~\ref{fig:high_level_structure}. 
Each image row is encapsulated as a packet, and these packets are reassembled to reconstruct the image. 
Previous research~\citep{jiang2023glitchhiker} has demonstrated that injected malicious signals can induce packet loss, leading to incorrect color interpretation, and consequently, color strips in the reconstructed image.

\begin{figure}[!t]
\centering
\includegraphics[width=1\columnwidth]{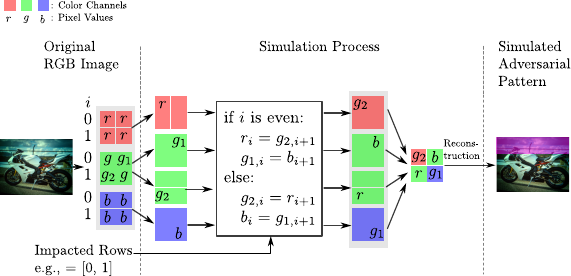}
\caption{The simulation process generates arbitrary adversarial patterns, i.e., color strips, in an RGB image.}
\label{fig:simulation_process}
\end{figure}

By carefully analyzing the attack mechanism explained in previous work, we observe that incorrect color interpretation can be characterized by a swap of red/blue and green channel pixel values.
We then propose a simulation method, outlined in Figure~\ref{fig:simulation_process}, to mimic this behavior.

The input of our method is an RGB image, with pixel values represented as $r$, $g$, and $b$ for red (R), green (G), and blue (B) channels, respectively. To simplify, we visualize the simulation process with two specific rows (two columns for each), termed ``Impacted Rows'' (row number, $i = 0, 1$). In particular, we simulate the computation based on the values of the upper-left corner in the r channel (denoted as $r$), the bottom-right corner of the blue channel (denoted as $b$), and the upper-right corner ($g_1$) and bottom-left corner ($g_2$) for green channel respectively.

Assuming neighboring pixel values are similar, we approximated color information using these four values, i.e., $r$, $g_{1}$, $g_{2}$, $b$.
For each impacted row with index $i$, if the row index $i$ is even, the red pixel value ($r$) is replaced with the green pixel value from the next row ($g_{2}$), and the green pixel value ($g_{1}$) is replaced with the blue pixel value from the next row ($b$).
If the row index $i$ is odd, the green pixel value ($g_{2}$) is replaced with the red pixel value from the next row ($r$), and the blue pixel value ($b$) is replaced with the green pixel value from the next row ($g_{1}$).

After applying the above transformations to the color channels, the pixels are reassembled to form a new image.
The reconstruction step, which is essentially demosaicing that reconstructs color information for all three channels, brings together the modified red, green, and blue channel values to form a new image.
The resulting image exhibits an adversarial pattern, as exemplified in the Figure~\ref{fig:simulation_process} where rows at the top (e.g., 0 and 1) are impacted. 
The position of the color strip can be controlled by adjusting the selected impacted rows.

\subsection{Evaluation of Similarity between Simulated Attack and Real Attack} \label{sec:real_simulate_evaluation}

\citet{zhang2024modeling} generated a set of real attack images by re-taking 100 randomly selected images from the COCO2017~\citep{lin2014microsoft} testing dataset using a physical setup to inject electromagnetic signals during the re-taking process. By adjusting the power of the injected signals, they conducted attacks at different severity levels, producing images under mild, moderate, and severe attacks. This process yields 300 under-attack images in total in their dataset, with three per original image, respective to the three severity levels. 
According to this real attack image set, the counts of color strips induced by mild, moderate, and severe attacks, fall in the ranges of \textit{[1, 6]}, \textit{[7, 12]}, and \textit{[13, 20]}, respectively. 

\begin{figure}[!t]
    \small
    \centering
    \begin{subfigure}[b]{0.48\columnwidth}
        \includegraphics[width=\textwidth]{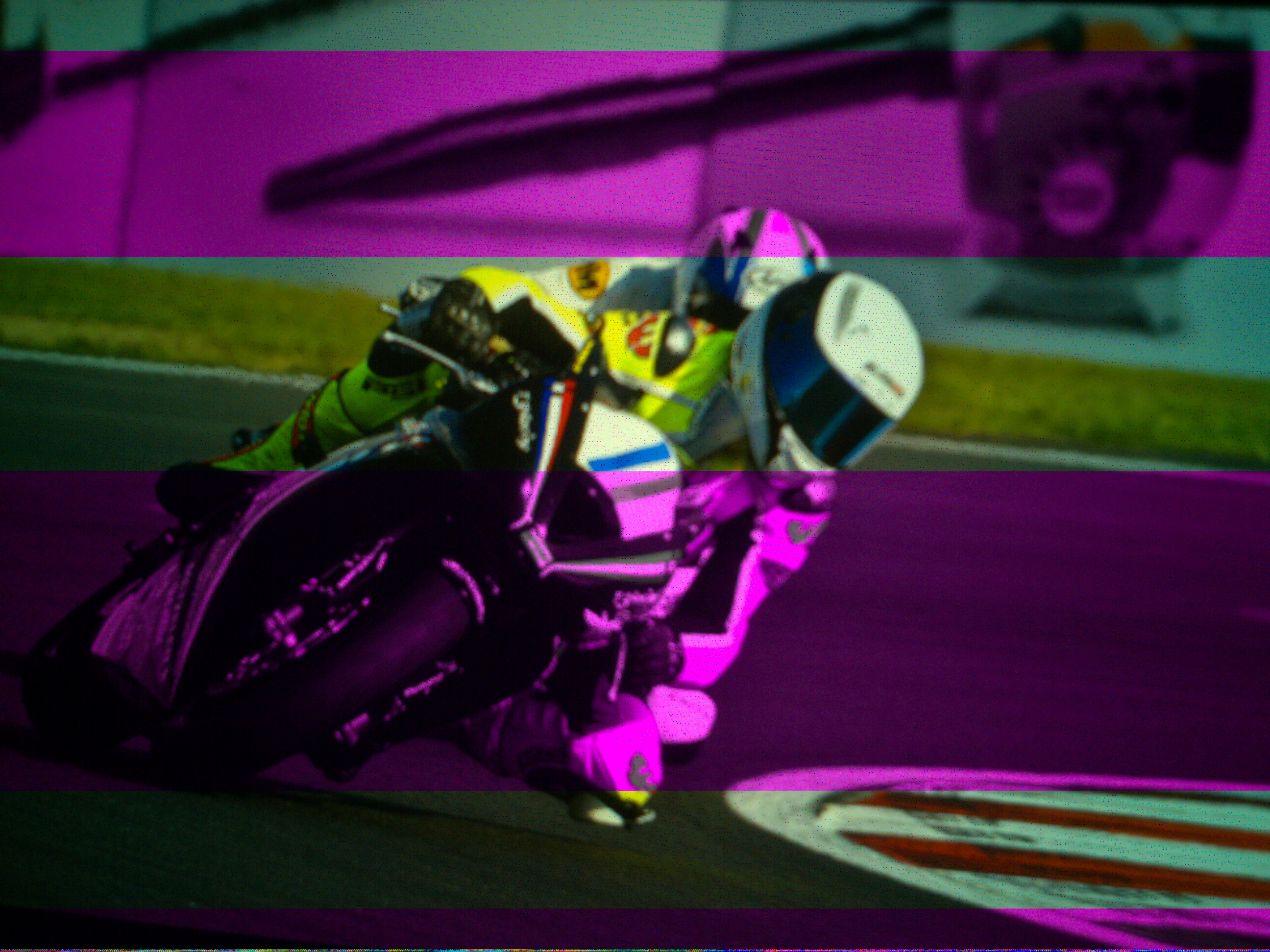}
    \end{subfigure}
    \hfill
    \begin{subfigure}[b]{0.48\columnwidth}
        \includegraphics[width=\textwidth]{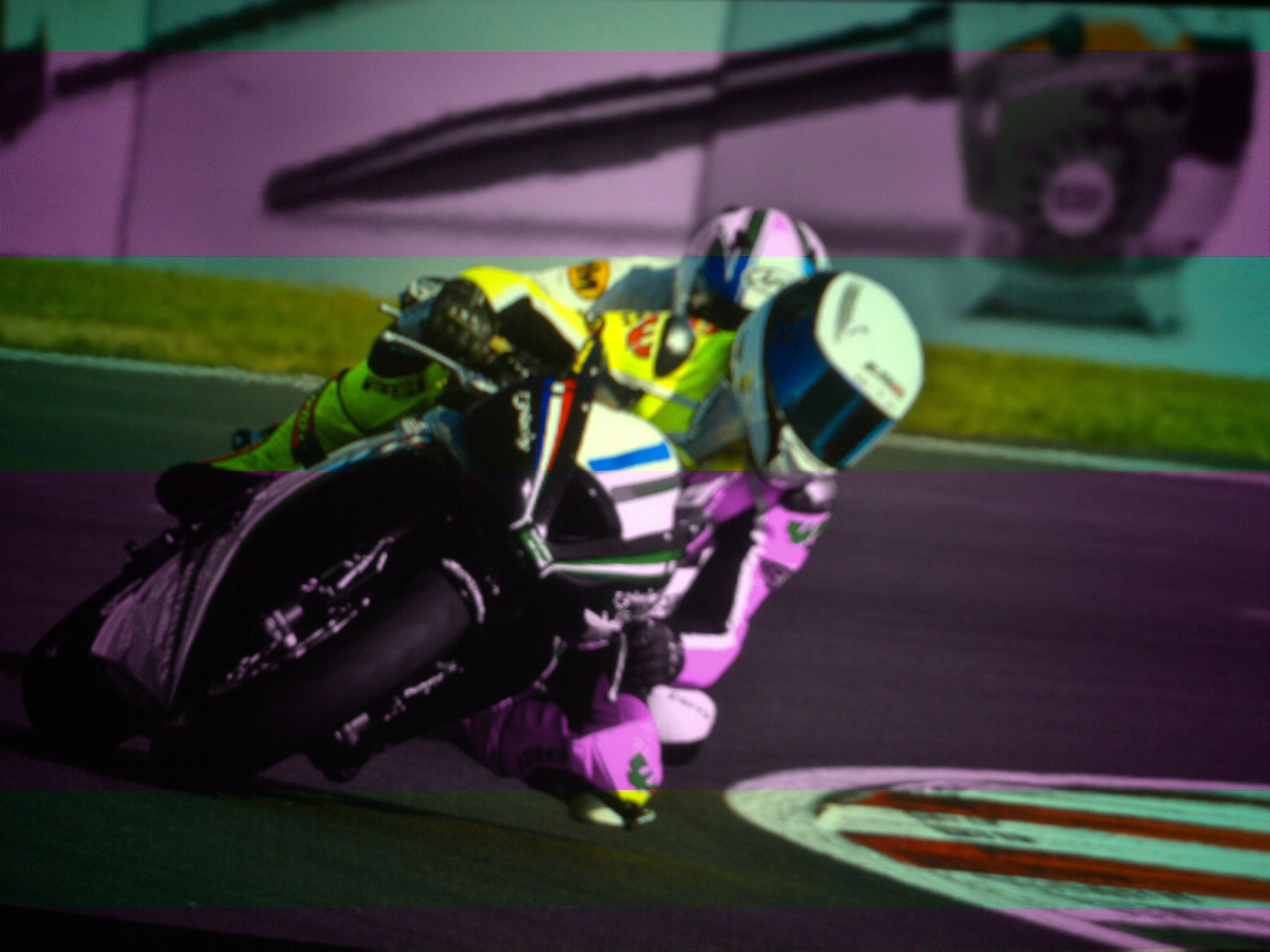}
    \end{subfigure}
    \centering \\
    \text{(a)} Mild attack power yields an average of 3 color strips

    \begin{subfigure}[b]{0.48\columnwidth}
        \includegraphics[width=\textwidth]{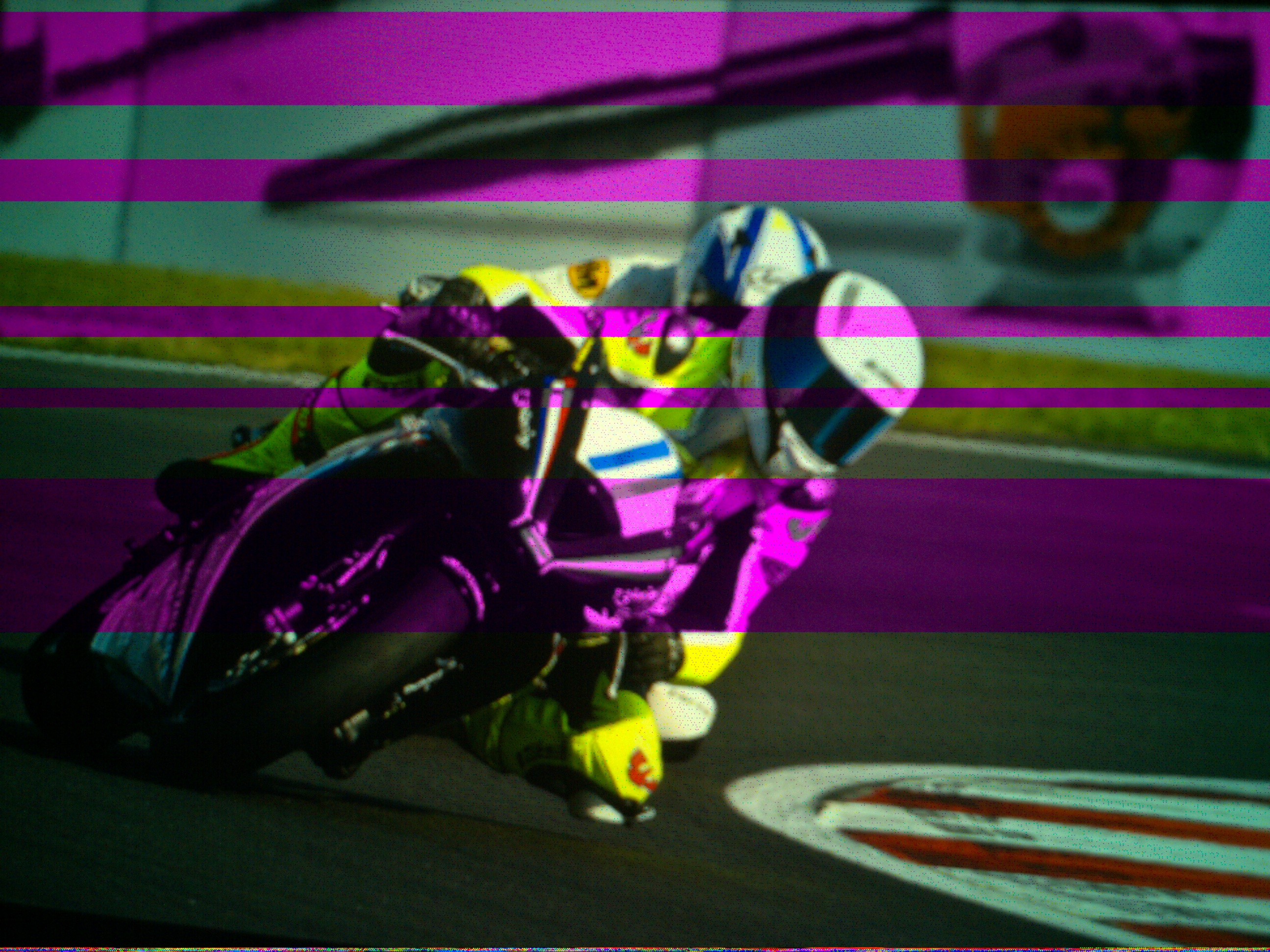}
    \end{subfigure}
    \hfill
    \begin{subfigure}[b]{0.48\columnwidth}
        \includegraphics[width=\textwidth]{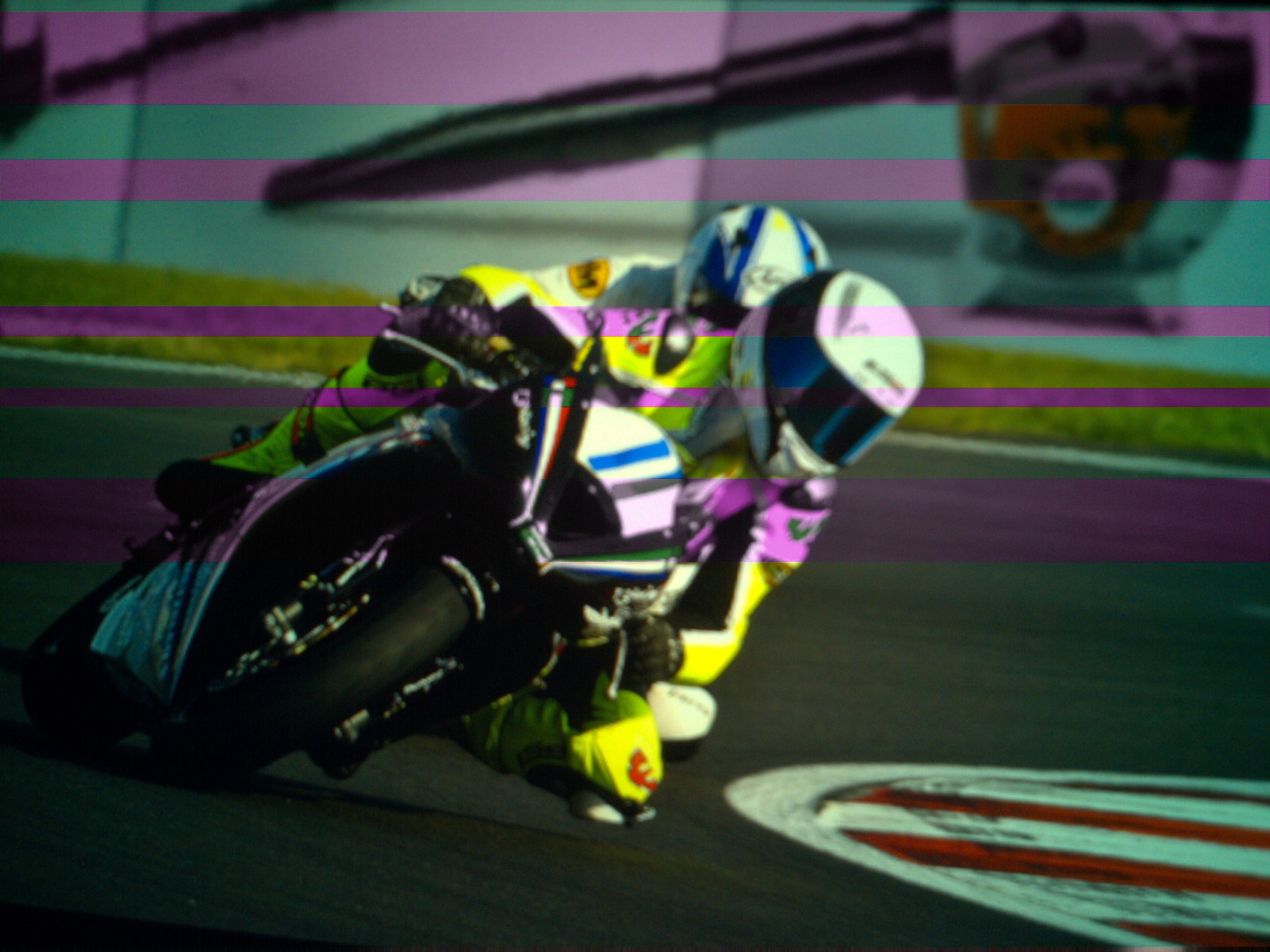}
    \end{subfigure}
    \centering \\
    \text{(b)} Moderate attack power yields an average of 6 color strips

    \begin{subfigure}[b]{0.48\columnwidth}
        \includegraphics[width=\textwidth]{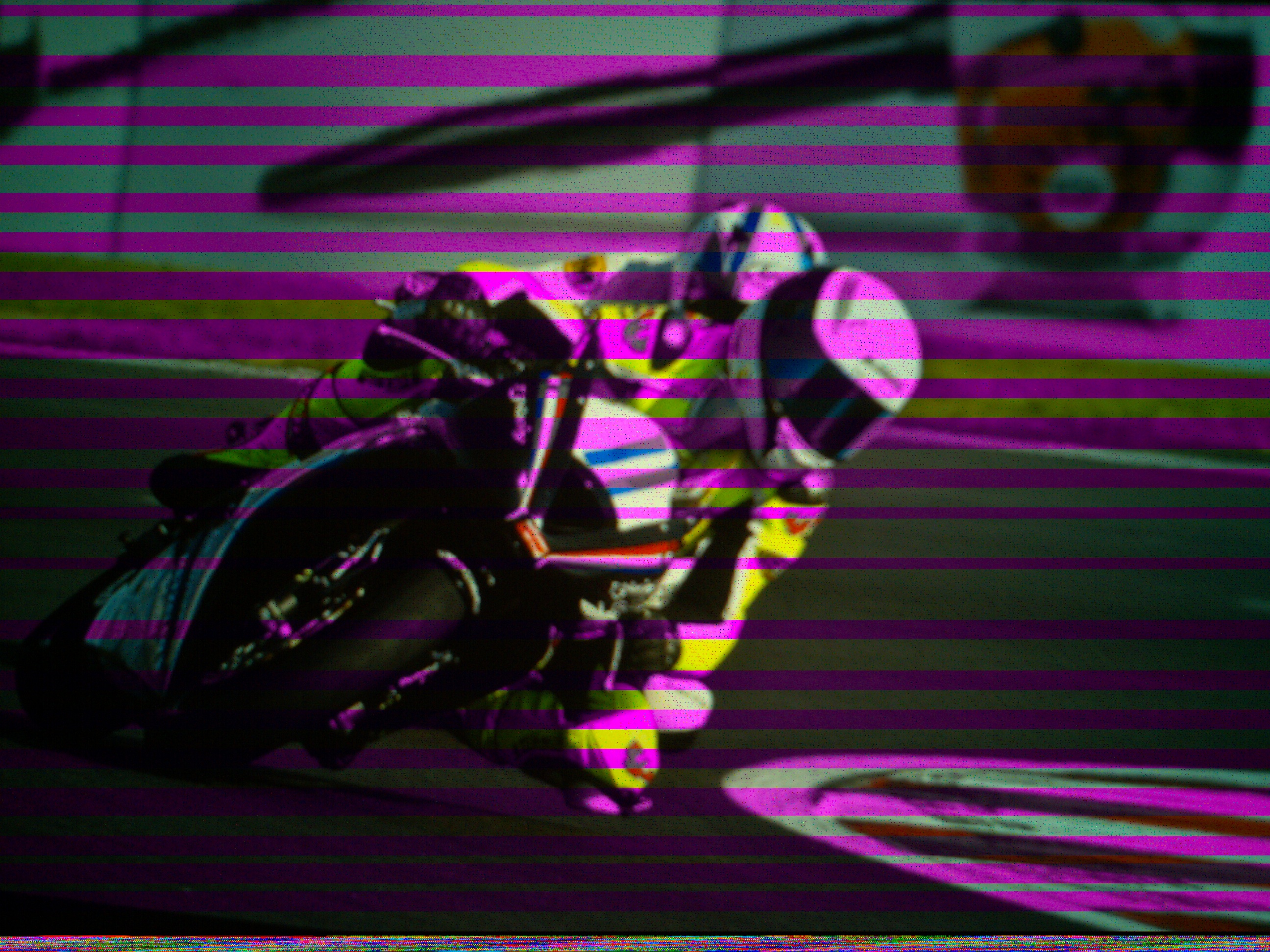}
    \end{subfigure}
    \hfill
    \begin{subfigure}[b]{0.48\columnwidth}
        \includegraphics[width=\textwidth]{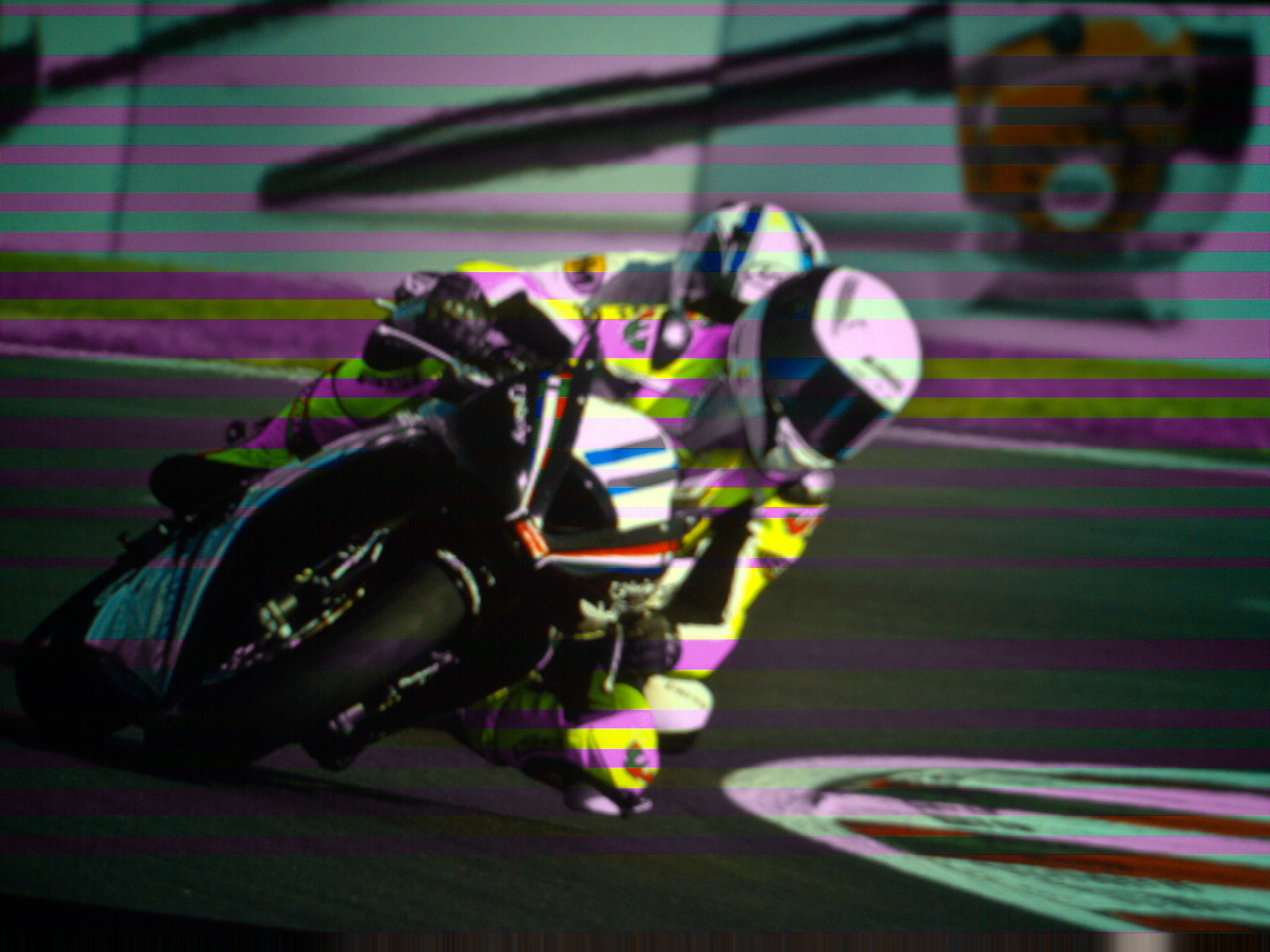}
    \end{subfigure}
    \centering \\
    \text{(c)} Severe attack power yields an average of 15 color strips

    \caption{Example real (left) and simulated (right) attack images with different levels of attack severity.}
    \label{fig:glitch_patterns}
\end{figure}

To validate the efficiency of our proposed ESIA simulation method. We apply our simulation method to the same 100 images, generating another set of 300 under-attack images, again, three per original image, one for each severity level. Specifically, we use the simulated attack method to produce the same number and location of color strips as the real ones (for each image at each severity level) as shown in Figure~\ref{fig:glitch_patterns}. We then evaluate the consistency of the impact of attacks -- the performance degradation they caused, between the images generated by our simulated method and those from real attacks in object detection task, one of the critical traffic-related applications (i.e., traffic object detection).

A diverse range of object detection models are selected for the evaluation, to ensure the comprehensiveness and reliability of the results. They include \textit{CNN-based single-stage detectors} (e.g., YOLOv3 \citep{redmon2018yolov3}, YOLOx \citep{ge2021yolox}, Retinanet \citep{lin2017retinanet}, VFNet \citep{zhang2021vfnet}, and EfficientDet \citep{tan2020efficientdet}), \textit{CNN-based multi-stage detectors} (e.g., Mask R-CNN \citep{he2017mask} and Cascade Mask R-CNN \citep{cai2019cascade}), and \textit{Transformer-based detectors} (e.g., DETR \citep{carion2020detr}, DINO \citep{zhang2022dino}, Co-DETR \citep{zong2023codetrs}, ViT-DET \citep{li2022vitdet}, and SWIN \citep{liu2021swin}).

\begin{figure}[!t]
\centering
\includegraphics[width=1\columnwidth]{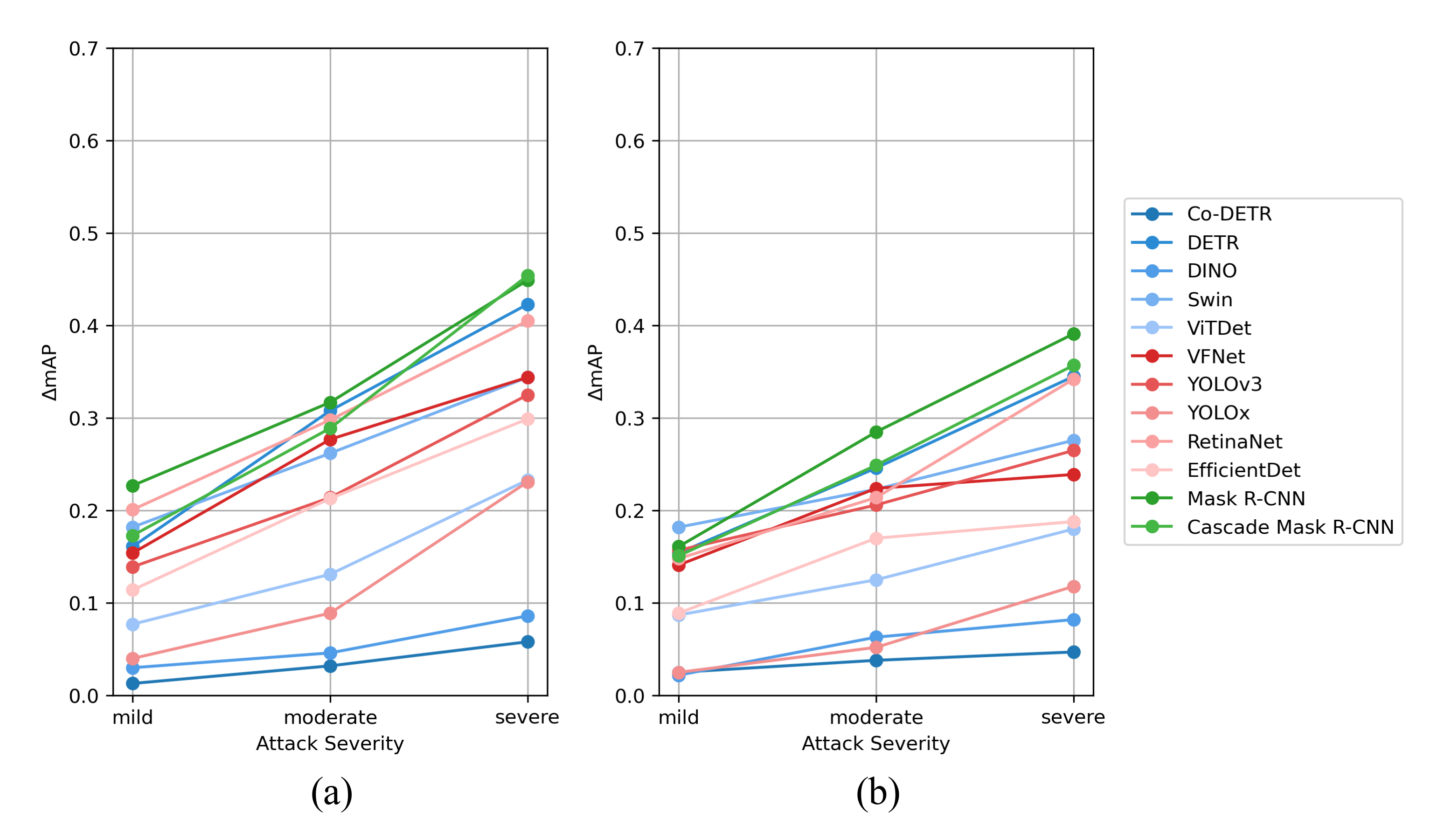} 

\caption{Similar performance ($\Delta$mAP50) between (a) real attack images and (b) simulated attack images across different object detection models.}
\label{similarity}
\end{figure}

We calculated the mean Average Precision (mAP) for each model based on the aforementioned data to illustrate the performance differences between simulated attack images and real attack images. Specifically, mAP50 is computed as the average of the Average Precision (AP) across all classes at a single Intersection over Union (IoU) threshold of 0.5. To visually demonstrate the performance degradation across models, we first calculated the mAP values (mAP50, mAP75, and mAP50:95) for each model on unattacked images, real attack images, and simulated attack images. Subsequently, we computed the differences in mAP values ($\Delta$mAP50, $\Delta$mAP75, and $\Delta$mAP50:95) between the attacked and unattacked states for both real and simulated attacks. The results for $\Delta$mAP50 are shown in Figure \ref{similarity}, where (a) represents performance degradation under real attacks and (b) illustrates degradation under simulated attacks. By examining the trend lines in the two image sets, it is evident that as the severity of the attack increases, the $\Delta$mAP values rise, indicating a decrease in mAP values. Moreover, the trend of $\Delta$mAP values under both real and simulated attacks is consistent across different models.

Furthermore, to evaluate whether there is a significant difference between the impacts of simulated and real attacks on AI models, we performed a t-test at a standard significance level of 5\% based on the calculated and summarized $\Delta$mAP results for all selected models under both attack types. The p-values for $\Delta$mAP50, $\Delta$mAP75, and $\Delta$mAP50:95 are shown in Table 1. As indicated by the results, all p-values from the t-tests for $\Delta$mAP exceed the 5\% threshold, indicating that there is insufficient evidence to suggest significant differences between the two datasets. These tests and analyses demonstrate that images generated by simulated attacks do not differ significantly from those generated by real attacks, and the impact on AI models follows a highly consistent trend.

\begin{table}[t]
\centering
  \begin{tabular}{cccc}
    \hline
     \textbf{Severity} & \textbf{$\Delta$mAP50} & \textbf{$\Delta$mAP75} & \textbf{$\Delta$mAP50:95} \\
    \hline
    Mild & 0.876 & 0.803 & 0.950 \\
    Moderate & 0.493 & 0.675 & 0.704 \\
    Severe & 0.491 & 0.373 & 0.896 \\
    \hline
  \end{tabular}
\caption{P-values of t-tests for different levels of $\Delta$mAP}
\label{table_map}
\end{table}

\subsection{Traffic Images with Simulated Adversarial Pattern}

We then apply the simulation method on a set of traffic images, to simulate attacks that autonomous vehicles might encounter in real world driving with complex environmental conditions. Specifically, we randomly select 10,000 images from the validation set of the BDD100k dataset~\citep{Yu_2020_CVPR}, which are well-annotated for traffic object detection and drivable area segmentation tasks.

We divide the images into three environmental groups with several subcategories based on driving conditions, specifically focusing on weather, scene, and time of day, -- three key driving environment variables.
To explore the impact of attacks in specific scenarios and ensure the reliability of the experimental results with a sufficient amount of data, we filter out undefined subcategories and some subcategories with fewer images. The number of images for each subcategory after filtering is shown in Table \ref{tab:image_distribution}.

To investigate the threat of ESIA at different severity levels, 
we randomly divide the images in each subcategory into four equal attack severity groups: unattacked, mild, moderate, and severe. Following the observations in the real attack dataset~\citep{zhang2024modeling}, for the mild, moderate, and severe groups, we perform simulated attacks with color strip numbers ranging from \textit{[1, 6]}, \textit{[7, 12]}, and \textit{[13, 20]}, respectively.

\begin{table}[t]
\small
\centering
\begin{tabular}{l|l|l}
\hline
\textbf{Weather} & \textbf{Time of Day} & \textbf{Scene} \\
\hline
\begin{tabular}[l]{@{}l@{}} Overcast (1239) \\ Clear (5346) \\ Rainy (738) \\ Snowy (769) \\ Partly Cloudy (738) \end{tabular} & 
\begin{tabular}[l]{@{}l@{}} Daytime (5258) \\ Night (3929) \\ Dawn (778) \end{tabular} & 
\begin{tabular}[l]{@{}l@{}} City Street (6112) \\ Highway (2499) \\ Residential (1253) \end{tabular} \\
\hline
Total (8830) & Total (9965) & Total (9864) \\
\hline
\end{tabular}
\caption{Distribution of images for three driving environmental groups (The numbers in parentheses indicate the number of images in each category).}
\label{tab:image_distribution}
\end{table}

\begin{table*}[t]

\small
\centering
\setlength\tabcolsep{5pt} 
\begin{tabular}{c|c|c|cccc|cccc}

\hline
\multirow{2}{*}{\textbf{Group}} & \multirow{2}{*}{\textbf{Sub-Category}}& \multirow{2}{*}{\textbf{Model}}& \multicolumn{4}{c}{\textbf{Traffic object detection (mAP50)}}& \multicolumn{4}{c}{\textbf{Drivable area segmentation (mIoU)}}\\

\cline{4-7}  \cline{8-11}

& &  & \textbf{No attack} &  \textbf{Mild} &  \textbf{Moderate}  & \textbf{Severe}   & \textbf{No attack} &  \textbf{Mild} &  \textbf{Moderate}  & \textbf{Severe}  
\\
\hline
\multirow{15}{*}{Weather} & \multirow{3}{*}{Clear} & HybridNets & 0.755 & 0.647 & 0.439 & 0.250& 0.908 & 0.889 & 0.842 & 0.790\\
&& A-YOLOM&  0.796 & 0.696 & 0.474 & 0.291& 0.914 & 0.892 & 0.843 & 0.791\\
&& YOLOP&  0.755 & 0.648 & 0.438 & 0.255& 0.913 & 0.898 & 0.856 & 0.807\\
\cline{2-11}

&\multirow{3}{*}{Rainy} & HybridNets & 0.800 & 0.695 & 0.501 & 0.341&  0.866 & 0.864 & 0.827 & 0.787\\
&& A-YOLOM & 0.835 & 0.746 & 0.538 & 0.378&  0.874 & 0.860 & 0.824 & 0.758\\
&& YOLOP & 0.782 & 0.680 & 0.486 & 0.337&   0.881 & 0.870 & 0.827 & 0.789\\
\cline{2-11}

&\multirow{3}{*}{Snowy} & HybridNets & 0.777 & 0.680 & 0.575 & 0.284 & 0.901 & 0.870 & 0.822 & 0.785\\
&& A-YOLOM & 0.808 & 0.733 & 0.517 & 0.322& 0.886 & 0.863 & 0.812 & 0.759 \\
&& YOLOP & 0.761 & 0.667 & 0.467 & 0.284& 0.907 & 0.878 & 0.841 & 0.789  \\
\cline{2-11}

&\multirow{3}{*}{Partly Cloudy} & HybridNets & 0.782 & 0.665 & 0.414 & 0.245&  0.906 & 0.880 & 0.842 & 0.790\\
&& A-YOLOM & 0.813 & 0.706 & 0.456 & 0.291& 0.919 & 0.887 & 0.827 & 0.781 \\
&& YOLOP & 0.768 & 0.662 & 0.421 & 0.257& 0.911 & 0.891 & 0.858 & 0.809 \\
\cline{2-11}

&\multirow{3}{*}{Overcast} & HybridNets & 0.785 & 0.702 & 0.465 & 0.268& 0.904 & 0.884 & 0.842 & 0.790 \\
&& A-YOLOM & 0.821 & 0.742 & 0.498 & 0.310& 0.909 & 0.895 & 0.846 & 0.800 \\
&& YOLOP & 0.780 & 0.697 & 0.458 & 0.277& 0.916 & 0.902 & 0.863 & 0.811 \\

\hline

\multirow{9}{*}{Scene} & \multirow{3}{*}{Highway} & HybridNets & 0.741 & 0.576 & 0.347 & 0.208&  0.915 & 0.894 & 0.846 & 0.790\\
&& A-YOLOM & 0.782 & 0.622 & 0.379 & 0.240& 0.928 & 0.897 & 0.848 & 0.791 \\
&& YOLOP & 0.742 & 0.575 & 0.347 & 0.212 & 0.917 & 0.895 & 0.854 & 0.800\\
\cline{2-11}

& \multirow{3}{*}{City Street} & HybridNets & 0.775 & 0.692 & 0.495 & 0.305&  0.897 & 0.875 & 0.834 & 0.789\\
&& A-YOLOM & 0.812 & 0.739 & 0.530 & 0.348&  0.902 & 0.880 & 0.834 & 0.780  \\
&& YOLOP & 0.769 & 0.683 & 0.489 & 0.307& 0.915 & 0.896 & 0.852 & 0.805 \\
\cline{2-11}

& \multirow{3}{*}{Residential} & HybridNets & 0.804 & 0.701 & 0.492 & 0.302&  0.907 & 0.895 & 0.840 & 0.795 \\
&& A-YOLOM & 0.837 & 0.748 & 0.541 & 0.349& 0.919 & 0.901 & 0.844 & 0.791 \\
&& YOLOP & 0.798 & 0.703 & 0.486 & 0.303& 0.913 & 0.893 & 0.849 & 0.796 \\

\hline

\multirow{9}{*}{Time of Day}&\multirow{3}{*}{Night} & HybridNets & 0.740 & 0.630 & 0.453 & 0.262&  0.903 & 0.884 & 0.839 & 0.796\\
&& A-YOLOM & 0.787 & 0.676 & 0.487 & 0.299&  0.910 & 0.888 & 0.842 & 0.794\\
&& YOLOP & 0.745 & 0.632 & 0.445 & 0.264&  0.908 & 0.894 & 0.855 & 0.811  \\
\cline{2-11}

&\multirow{3}{*}{Dawn} & HybridNets & 0.787 & 0.716 & 0.440 & 0.263& 0.897 & 0.883 & 0.835 & 0.792\\
&& A-YOLOM & 0.824 & 0.756 & 0.469 & 0.304& 0.894 & 0.888 & 0.844 & 0.780  \\
&& YOLOP & 0.777 & 0.718 & 0.428 & 0.267& 0.907 & 0.896 & 0.843 & 0.800  \\
\cline{2-11}

&\multirow{3}{*}{Daytime} & HybridNets & 0.796 & 0.686 & 0.473 & 0.283&  0.906 & 0.884 & 0.836 & 0.790\\
&& A-YOLOM &  0.829 & 0.727 & 0.513 & 0.327& 0.907 & 0.884 & 0.834 & 0.782\\
&& YOLOP & 0.783 & 0.680 & 0.465 & 0.289&  0.917 & 0.896 & 0.852 & 0.801\\

\bottomrule

\end{tabular}

\caption{Vulnerability of models to different environmental conditions in traffic object detection and drivable area
segmentation. }
\label{table3}

\end{table*}

\begin{table*}[t]
\small
\centering
\setlength\tabcolsep{5pt} 

\begin{tabular}{c|c|ccc|ccc}
\hline
\multirow{2}{*}{\textbf{Group}}& \multirow{2}{*}{\textbf{Sub-Category}}&\multicolumn{3}{c}{\textbf{Threat on traffic object detection}}& \multicolumn{3}{c}{\textbf{Threat on drivable area segmentation}}\\
\cline{3-5} \cline{6-8}
&   & \textbf{$D\_S_{\text{mild}}$} &  \textbf{$D\_S_{\text{moderate}}$} &  \textbf{$D\_S_{\text{severe}}$}  & \textbf{$D\_S_{\text{mild}}$} &  \textbf{$D\_S_{\text{moderate}}$} &  \textbf{$D\_S_{\text{severe}}$}       
\\
\hline
\multirow{5}{*}{Weather} & Clear &-13.68\%	&-41.43\%	&-65.52\% &-2.05\%	&-7.09\%	&-12.69\%\\
&Rainy &-12.28\%	&-36.93\%	&-56.34\% &-1.03\%	&-5.45\%	&-10.95\%\\
& Snowy &-11.37\%	&-33.55\%	&-62.09\%  &-3.08\%	&-8.13\%	&-13.41\%\\
&Partly Cloudy &-13.97\%	&-45.38\%	&-66.47\% &-2.85\%	&-7.63\%	&-13.01\%\\
&Overcast &-10.28\%	&-40.46\%	&-64.20\% &-1.76\%	&-6.53\%	&-12.02\%\\
\hline

\multirow{3}{*}{Scene} & Highway & -21.74\%	&-52.65\%	&-70.89\% & -2.68\%	&-7.68\%	&-13.73\%\\
& City Street &-10.29\%	&-35.76\%	&-59.29\% &-2.32\%	&-7.15\%	&-12.53\% \\
& Residential &-11.78\%	&-37.76\%	&-60.92\% &-1.82\%	&-7.52\%	&-13.03\%\\
\hline

\multirow{3}{*}{Time of Day} & Night &-14.71\%	&-39.06\%	&-63.72\% &-2.02\%	&-6.80\%	&-11.76\%\\
& Dawn &  -8.29\%	&-44.03\%	&-65.11\% & -1.15\%	&-6.52\%	&-12.08\%\\
& Daytime & -13.09\%	&-39.77\%	&-62.70\% & -2.42\%	&-7.62\%	&-13.08\%\\

\bottomrule
\end{tabular}

\caption{The threat of simulated attack on traffic object detection (i.e., degradation in detection accuracy in mAP) and drivable area segmentation (i.e., degradation of segmentation accuracy in mIoU) across different driving environmental conditions.}
\label{table4}
\end{table*}
\begin{table*}[t]
\small
\centering
\setlength\tabcolsep{5pt} 

\begin{tabular}{c|c|ccc|ccc}
\hline
\multirow{2}{*}{\textbf{Model}}& \multirow{2}{*}{\textbf{Group}}&\multicolumn{3}{c}{\textbf{Threat on traffic object detection}}& \multicolumn{3}{c}{\textbf{Threat on drivable area segmentation}}\\
\cline{3-5} \cline{6-8}
& & \textbf{$D\_M_{\text{mild}}$} &  \textbf{$D\_M_{\text{moderate}}$} &  \textbf{$D\_M_{\text{severe}}$}     & \textbf{$D\_M_{\text{mild}}$} &  \textbf{$D\_M_{\text{moderate}}$} &  \textbf{$D\_M_{\text{severe}}$}
\\
\hline
\multirow{3}{*}{HybridNets} & Weather &-13.09\%	&-38.61\%	&-64.45\%&-2.17\%	&-6.89\%	&-12.08\%\\
& Scene & -15.26\%	&-42.70\%	&-65.00\% & -2.02\%	&-7.32\%	&-12.68\%\\
& Time of Day & -12.57\%	&-41.15\%	&-65.21\% & -2.03\%	&-7.24\%	&-12.12\%\\
\hline

\multirow{3}{*}{A-YOLOM} & Weather &-11.06\%	&-39.06\%	&-60.95\% &-2.33\%	&-7.76\%	&-13.61\%\\
& Scene & -13.36\%	&-40.54\%	&-61.59\%& -2.58\%	&-8.11\%	&-14.07\%\\
& Time of Day & -11.55\%	&-39.77\%	&-61.89\%& -1.87\%	&-7.04\%	&-13.09\%\\
\hline

\multirow{3}{*}{YOLOP} & Weather &-12.80\%	&-40.99\%	&-63.37\%&-1.96\%	&-6.25\%	&-11.54\%\\
& Scene &  -15.20\%	&-42.91\%	&-64.51\%&  -2.22\%	&-6.92\%	&-12.53\%\\
& Time of Day &-11.97\%	&-41.93\%	&-64.43\%&-1.68\%	&-6.66\%	&-11.71\%\\

\bottomrule
\end{tabular}

\caption{The threat of simulated attack on traffic object detection (i.e., degradation in detection accuracy in mAP) and drivable area segmentation (i.e., degradation of segmentation accuracy in mIoU) across different models.}
\label{table5}
\end{table*}

\section{Impacts of ESIA in Different Driving Scenarios}

This section will explore and discuss the impacts of ESIA on perception tasks across various driving scenarios.

\subsection{Tasks and Models}
We select two critical tasks, which are object detection and drivable area segmentation~\citep{feng2020deep}. Object detection is an important task to identify a specific object in an image and determine its location. 
The detection not only needs to identify the categories of objects in the image but also needs to mark the exact position of each object in the image.
Drivable area segmentation is one of the key tasks in autonomous driving designed to identify and segment drivable areas in road scenarios. The goal of this task is to detect areas where the vehicle is safe to travel in order to aid in path planning, obstacle avoidance, and decision-making for autonomous vehicles.

Considering that multi-task models offer better resource efficiency compared to single-task models and are more practical for resource-limited devices, we select three state-of-the-art multi-task models: HybridNets~\citep{vu2022hybridnets}, A-YOLOM~\citep{wang2024you}, and YOLOP~\citep{wu2022yolop}. Specifically, HybridNets is based on EfficientNet~\citep{tan2019efficientnet}, a CNN architecture known for delivering comparable performance to other CNN architectures with fewer resources. A-YOLOM, based on YOLOv8, excels in detection and segmentation within the real-time-oriented YOLO family. YOLOP utilizes a Spatial Pyramid Pooling (SPP)~\citep{he2015spatial} module for feature generation and fusion. Previous studies~\citep{zhang2021evaluating,ranjan2019attacking} have demonstrated that the SPP module is robust against adversarial attacks, and we wonder whether it also demonstrates robustness in the context of ESIA.

For evaluation metrics, we follow standard practices: we measure mAP50 for traffic object detection and mIoU for drivable area segmentation. Average Precision is computed as the area under the precision-recall curve. The Intersection over Union (IoU) metric is used to evaluate the drivable area, and mIoU is the average IoU for each class. To ensure a fair comparison between different models, we follow the evaluation setup specified in the work of three selected models, which means a confidence threshold of 0.001 and a Non-Maximum Suppression (NMS) threshold of 0.6 for all three models. We use the best-performing pretrained models of HybridNets, A-YOLOM, and YOLOP on the BDD100k dataset.  It is worth noting that all three pre-trained models focus exclusively on four classes -- car, bus, truck, and train during training, and combine these classes into a single ``vehicle'' classification for the detection task.

\subsection{Experimental Results and Analysis}
In the experiments, driving scenarios are defined by three categories: weather, scene, and time of day, with subcategories such as residential, city street, and highway for the scene, as detailed in Table \ref{tab:image_distribution}. The results for each category are presented in Table \ref{table3}. It can be observed that model performance significantly deteriorates with increasing attack intensity across all categories. For example, in the HybridNets model under the clear subcategory, the mAP50 for detection decreases by 10.80\%, 31.60\%, and 50.50\% under mild, moderate, and severe intensity attacks, respectively, compared to the unattacked condition. Correspondingly, the mIoU for segmentation drops by 1.9\%, 6.6\%, and 11.8\% under the same conditions.

To clearly illustrate the impact of attack across different driving scenarios and models, we further calculate the percentage decrease in performance compared to the ``No Attack'' condition, referred to as Degradation. Furthermore, We calculate $D\_S_i$ by averaging the Degradation across the three models for the same attack intensity within the same subcategory and $D\_M_i$ by averaging the Degradation across all subcategories for the same attack intensity within the same model, where $_i$ represents attack intensity, i.e., mild, moderate, severe. The results are presented in Table \ref{table4} and Table \ref{table5}, respectively. 

It is observed that: (1) For the weather category, in the traffic object detection task, all three models are less affected by simulated attacks in the snowy subcategory compared to other subcategories. Conversely, in the drivable area segmentation task, the models experience less impact from attacks in the rainy subcategory. 
(2) For scene types, regardless of whether the task is traffic object detection or drivable area segmentation, all three models tend to experience the least impact from attacks in the city street subcategory. This may be attributed to the more complex and cluttered nature of city streets, which compels the models to focus on finer details, thereby making them less susceptible to attacks. In contrast, highway scenarios, which typically feature simpler and more predictable environments, are most affected by attacks. This observation highlights how the simplicity of an environment can exacerbate the impact of ESIA on model performance.
(3) For model robustness, all three models are significantly affected. In the comparison among models, A-YOLOM experiences the least impact in traffic object detection, while YOLOP is the least affected in drivable area segmentation.

\subsection{Potential Driving Risks}

\begin{figure}[!t]
\centering
\includegraphics[width=1\columnwidth]{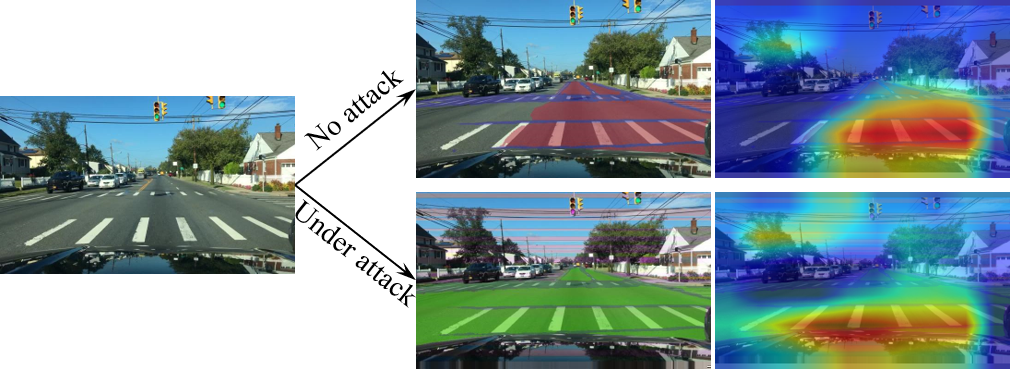} 

\caption{Model attention variations across attack intensities in the case of ``driving against traffic”.}
\label{Driving against traffic}
\end{figure}

\begin{figure}[!t]
\centering
\includegraphics[width=1\columnwidth]{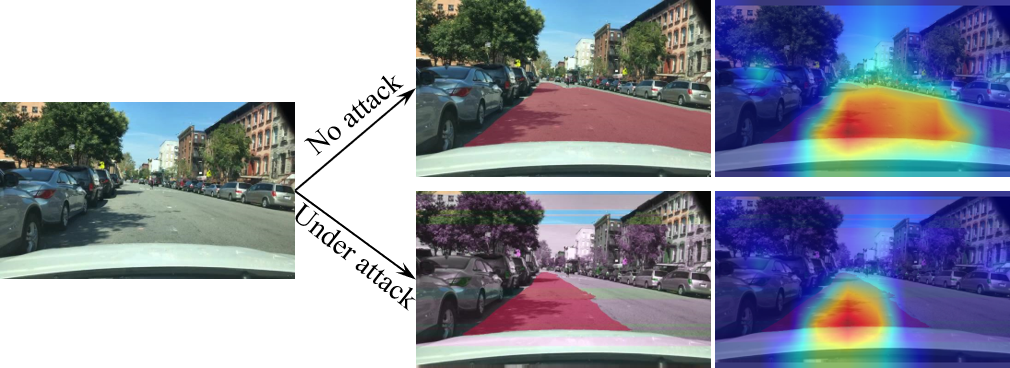} 

\caption{Model attention variations across attack intensities in the case of ``drivable area reduction”.}
\label{Drivable area reduction}
\end{figure}

The above experiments focus on changes in model metrics. We know that not every erroneous prediction by a model will lead to driving risks, though it may cause a decline in model metrics. Does a decline in model metrics pose risks in real-world driving scenarios? In the following parts, we discuss cases that could lead to driving risks.

We visualize the model prediction results for images in the ``no attack'' and ``under attack'' conditions, with prediction parameters consistent with evaluation parameters. For traffic object detection, previous studies have demonstrated issues like hiding and creating objects in the presence of attacks~\citep{jiang2023glitchhiker, zhang2024understanding}, and we do not repeat them here. 
In this part, we focus on cases related to drivable area segmentation. We identify two potential driving risk scenarios: (1) Driving against traffic: After the attack, the detection of reverse lanes changes from non-drivable to drivable. (2) Drivable area reduction: After the attack, the model detects a significant reduction in the drivable area.

To further investigate the changes in model attention related to the two potential driving risk scenarios caused by simulated attacks, we apply Grad-CAM~\citep{selvaraju2017grad} to the feature extraction layer (e.g., bifpn[5].conv6-down) close to the segmentation head. In Figure \ref{Driving against traffic}, it can be observed that after the attack, the model tends to focus more on erroneous areas, which indicates a shift in the model’s attention from relevant features to misleading ones, thereby compromising its performance and reliability. However, as shown in Figure \ref{Drivable area reduction}, the model’s focus on the correct regions decreases, indicating that the attack causes a significant reduction in attention towards relevant features, even at mild attack intensity. Interestingly, as the attack intensity increases, the model tends to revert to its pre-attack state. We speculate that the higher attack intensity affects more modules within the model simultaneously, leading to a counteracting effect that diminishes the overall impact of the attack. We will explore this phenomenon further in our future work.  

\section{Future Work}
In future work, we plan to extend our methodology to evaluate the impact of integrating data streams from multiple sensors (e.g., cameras, LiDARs, radars) on end-to-end models under electromagnetic signal injection attacks (ESIA). 
A key focus will be examining whether the compromise of camera systems introduces security vulnerabilities when multiple sensor modalities are present. Additionally, we aim to broaden our evaluations to include fully autonomous systems, assessing their resilience in real-world scenarios. 
Lastly, we will investigate mitigation strategies for ESIA, designing more robust and resilient systems.

\section{Conclusion}
In this paper, we systematically explore the previously unaddressed vulnerabilities of the perception system in autonomous vehicles to electromagnetic signal injection attacks (EISA) in various driving scenarios. 
An efficient method is developed for to simulate adversarial patterns, generating a set of under-attack traffic images for the analytics in this and future work. 
Our extensive experiments reveal that model performance degrades to varying extents under attacks, with our observations highlighting severe safety risks such as contraflow driving. This further underscores the significance of evaluating AI models' robustness and reliability in the face of potential adversarial attacks.

This study establishes a foundation for future investigation of techniques to enhance the robustness of AI models (in autonomous driving systems) against such attacks, thereby strengthening security and providing passengers with reliable and safe services. It also showcases a simulation and evaluation framework for other security- or safety-critical applications under potential attacks, shedding lights on future research on AI robustness.

\section{Acknowledgements}

This work was, in part, supported by the Hong Kong Polytechnic University under grant P0048514, and Shenzhen University.

\bibliography{aaai25}

\end{document}